\RequirePackage{ifpdf}
\ifpdf 
\documentclass[pdftex]{sigma}
\else
\documentclass{sigma}
\fi

\usepackage{mathrsfs}
\usepackage{bbm}
\usepackage{bm}
\usepackage[vcentermath,enableskew]{youngtab}

\newcommand{\unit}{\mathbbm{1}}   			

\newcommand{\CC}{\mathcal{C}}

\newcommand{\CCD}{\mathscr{D}}

\newcommand{\CCH}{\mathscr{H}}
\newcommand{\CI}{\mathcal{I}}
\newcommand{\CCI}{\mathscr{I}}

\newcommand{\CO}{\mathcal{O}}

\newcommand{\CP}{\mathcal{P}}

\newcommand{\CZ}{\mathcal{Z}}

\newcommand{\FR}{\mathbbm{R}}     			
\newcommand{\FC}{\mathbbm{C}}     			
\newcommand{\NN}{\mathbbm{N}}     			
\newcommand{\CPP}{{\mathbbm{C}P}}    			
\newcommand{\ah}{\hat{a}}

\newcommand{\fh}{\hat{f}}

\newcommand{\dd}{\mathrm{d}}     			
\newcommand{\dif}[1]{\mathrm{d}#1\,}     			
\newcommand{\diag}{{\mathrm{diag}}}     		
\newcommand{\de}{\mathrm{e}}     			
\newcommand{\di}{\mathrm{i}}     			
\newcommand{\eps}{{\varepsilon}}			

\newcommand{\bz}{{\bar{z}}}

\DeclareMathOperator{\tr}{tr}

\newcommand{\au}{\mathfrak{u}}
\newcommand{\asu}{\mathfrak{su}}

\newcommand{\sU}{\mathsf{U}}     			

\newcommand{\sSU}{\mathsf{SU}}

\newcommand{\sS}{\mathsf{S}}

\newcommand{\sEnd}{\mathsf{End}\,}

\def\tyng(#1){\hbox{\tiny $\yng(#1)$}}			
\def\tyoung(#1){\hbox{\tiny $\young(#1)$}}			

\begin{document}

\allowdisplaybreaks

\renewcommand{\thefootnote}{$\star$}

\renewcommand{\PaperNumber}{050}

\FirstPageHeading

\ShortArticleName{The Multitrace Matrix Model of Scalar Field Theory on Fuzzy $\CPP^n$}

\ArticleName{The Multitrace Matrix Model of\\ Scalar Field Theory on Fuzzy $\boldsymbol{\CPP}^{\boldsymbol{n}}$\footnote{This paper is a
contribution to the Special Issue ``Noncommutative Spaces and Fields''. The
full collection is available at
\href{http://www.emis.de/journals/SIGMA/noncommutative.html}{http://www.emis.de/journals/SIGMA/noncommutative.html}}}

\Author{Christian S{\"A}MANN~$^{\dag\ddag}$}

\AuthorNameForHeading{C. S{\"a}mann}

\Address{$^\dag$~Department of Mathematics, Heriot-Watt University,
Colin Maclaurin Building,\\
\hphantom{$^\dag$}~Riccarton, Edinburgh EH14 4AS, UK}
\EmailD{\href{mailto:C.Saemann@hw.ac.uk}{C.Saemann@hw.ac.uk}}
\URLaddressD{\url{http://www.christiansaemann.de}}

\Address{$^\ddag$~Maxwell Institute for Mathematical Sciences, Edinburgh, UK}

\ArticleDates{Received March 25, 2010, in f\/inal form June 03, 2010;  Published online June 11, 2010}

\Abstract{We perform a high-temperature expansion of scalar quantum f\/ield theory on fuzzy~$\CPP^n$ to third order in the inverse temperature. Using group theoretical methods, we rewrite the result as a multitrace matrix model. The partition function of this matrix model is evaluated via the saddle point method and the phase diagram is analyzed for various~$n$. Our results conf\/irm the f\/indings of a previous numerical study of this phase diagram for~$\CPP^1$.}


\Classification{81T75}

\renewcommand{\thefootnote}{\arabic{footnote}}
\setcounter{footnote}{0}

\section{Introduction}

Fuzzy spaces are noncommutative geometries which arise from quantizing certain compact K{\"a}hler manifolds. The most prominent such space is the fuzzy sphere, which was f\/irst constructed by Berezin \cite{Berezin:1974du}. In the original construction, the aim was the same as that of geometric quantization, i.e.\ to provide a general quantization prescription for a particle whose phase space is an arbitrary Poisson manifold. Today, fuzzy spaces attract most interest for dif\/ferent reasons: First, fuzzy spaces appear quite naturally in various contexts in string theory where they replace parts of the classical geometry of the target space with an approximate quantum geo\-met\-ry. Closely related is the observation that fuzzy geometries seem to emerge from the dynamics of matrix models and thus they could be crucial in background independent formulations of theo\-ries of gravity. And f\/inally one can regulate quantum f\/ield theories on K{\"a}hler manifolds by putting the theory on the corresponding Berezin-quantized or fuzzy manifold.

The idea of using fuzzy spaces as regulators for quantum f\/ield theories goes back to the early 1990's \cite{Madore:1991bw,Grosse:1995ar}. This approach is very appealing, as the def\/inition of scalar quantum f\/ield theories on fuzzy spaces is under complete control: All functional integrals are automatically well-def\/ined because the algebra of functions on a fuzzy space is f\/inite dimensional. Taking the large volume limit of the fuzzy space, we can even regulate scalar quantum f\/ield theories on f\/lat spaces and thus try to compete with the lattice approach. The main advantage of fuzzy regularization over the latter is that all the isometries of the original K{\"a}hler manifold survive the quantization procedure.

Particularly nice spaces to use in a fuzzy regularization are the complex projective spaces, as they are the Berezin-quantizable manifolds with the largest possible symmetry groups. Furthermore, their quantization is straightforward and can be done completely in terms of group theory. As usual in a ``good'' quantization, real functions are mapped to hermitian operators on a Hilbert space, which is f\/inite dimensional in Berezin quantization. Real scalar f\/ield theories are therefore simply hermitian matrix models.

The most prominent hermitian matrix models are given by a potential consisting of a trace over a polynomial in the matrix variable. One can therefore switch directly to an eigenvalue formulation. In the case of scalar f\/ield theories on fuzzy spaces, this is not possible because the kinetic term yields a coupling to a number of f\/ixed ``external'' matrices.

A f\/irst attempt at gaining an analytical handle on fuzzy scalar f\/ield theories was made in \cite{Steinacker:2005wj}. A new method to overcome the problem of external matrices was then proposed in
\cite{O'Connor:2007ea}. Here, a~high-temperature expansion of the kinetic term in the partition function was performed and the resulting expressions could be evaluated analytically via group theoretic methods. It was shown that the resulting partition function can be rewritten as the partition function of a multitrace matrix model. This partition function can then be computed analytically for both f\/inite and inf\/inite matrix sizes using, e.g., orthogonal polynomials or the saddle point approximation. For~$\CPP^1$, this computation was performed to second order in the inverse temperature $\beta$ in~\cite{O'Connor:2007ea}. In this paper, we continue this work and generalize the results to third order in $\beta$ and to arbitrary~$\CPP^n$.

One of the motivations for this work is to explain the phase diagram for scalar f\/ield theory on fuzzy $\CPP^1$ which has been obtained via numerical methods in~\cite{GarciaFlores:2005xc}, see also~\cite{Panero:2006bx} for a more detailed study as well as~\cite{Panero:2006cs} for a review and further numerical results. The numerical results suggest that the phase diagram is invariant under a particular multiscaling. We can therefore restrict ourselves to the limit of inf\/inite matrix size, in which we can use the saddle point approximation to compute the partition function of our model.

Further reasons to compute the multitrace matrix model to third order in $\beta$ are the possibility to use this result in a similar study of scalar f\/ield theory on $\FR\times \CPP^1$ as well as our intent to discuss the link to (deformed) integrable hierarchies in future work.

In the analysis of the phase diagram, we will focus our attention on the three lowest-dimensional fuzzy spaces $\CPP^1_F$, $\CPP^2_F$ and $\CPP^3_F$. In the f\/irst case, the goal will be to compare the resulting phase diagram with the numerically obtained one. The quantum f\/ield theory on the second space corresponds in the large volume limit to a scalar quantum f\/ield theory of $\phi^4$-type on $\FR^4$. While admittedly it is not clear what the Lagrangian of the f\/ield theory on $\FR^4$ being regularized actually is, this presents an example of both a well-def\/ined and renormalizable four-dimensional noncommutatively deformed $\phi^4$-theory. The theory on $\CPP^3$ could be interpreted as a regularization of a non-renormalizable f\/ield theory, and one might hope for signs of this in the matrix model.

The paper is structured as follows. In Section~\ref{section2}, we review the construction of fuzzy $\CPP^n$ and scalar f\/ield theory on this noncommutative space. Section~\ref{section3} describes the high-temperature expansion in detail and the results are presented to order $\beta^3$. In Section~\ref{section4}, we analyze the thus obtained multitrace matrix model for the three lowest-dimensional fuzzy $\CPP^n$ and we conclude in Section~\ref{section5}. Conventions, rather technical details and helpful intermediate results are given in the appendix.

\section[Scalar field theory on fuzzy CP**n]{Scalar f\/ield theory on fuzzy $\boldsymbol{\CPP^n}$}\label{section2}

The general mathematical framework containing the quantization of complex projective space which is referred to as {\em fuzzy $\CPP^n$} in the physics literature is known as {\em Berezin--Toeplitz quantization}, see e.g.\ \cite{IuliuLazaroiu:2008pk} and references therein for a detailed discussion. In the case of $\CPP^n$, there is a shortcut to the general constructions of Berezin--Toeplitz quantization which originates from the fact that $\CPP^n$ is the coset space $\sU(n+1)/\sU(1)\times\sU(n)$. We will use this group theoretic approach here, as it has the additional advantage of allowing for simple computations of quantities like spectra of quadratic Casimirs and their eigenspaces, which we will need for our further discussion.

\subsection[Berezin quantization of CP**n]{Berezin quantization of $\boldsymbol{\CPP^n}$}

The Hilbert space $\CCH_\ell$ which we use in Berezin quantizing $\CPP^n$ is the space of global holomorphic sections of the line bundle $\CO(\ell)$ over $\CPP^n$ with $\ell\geq 0$. As a vector space, $\CCH_\ell$ is spanned by the homogeneous polynomials of degree $\ell$ in the homogeneous coordinates $z^0,\ldots,z^n$ on $\CPP^n$. Recall that $\CPP^n\cong\sSU(n+1)/\sS(\sU(1)\times\sU(n))$, and $\CCH_\ell$ forms a representation of $\sSU(n+1)$ which is given by the totally symmetrized tensor product of $\ell$ fundamental representations. In terms of Dynkin labels, this representation reads as $(\ell,0,\ldots,0)$ and has dimension
\begin{equation*}
 N_{n,\ell}:=\dim(\CCH_\ell)=\dim(\ell,0,\ldots,0)=\frac{(n+\ell)!}{n!\ell!}.
\end{equation*}

We will f\/ind it convenient to map the polynomials to elements of the $\ell$-particle Hilbert space in the Fock space of $n+1$ harmonic oscillators with creation and annihilation operators satisfying the algebra $[\ah_\alpha,\ah_\beta^\dagger]=\delta_{\alpha\beta}$ and $\ah_\alpha |0\rangle=0$ for $\alpha,\beta=0,\ldots,n$. We thus identify
\begin{equation*}
 \CCH_\ell\cong{\rm span}(\ah^\dagger_{\alpha_1}\cdots\ah^\dagger_{\alpha_\ell}|0\rangle).
\end{equation*}
The Berezin symbol map $\sigma_\ell:\sEnd(\CCH_\ell)\rightarrow \CC^\infty(\CPP^n)$ is def\/ined as
\begin{equation*}
 \sigma_\ell(\fh)(z):=\langle z,\ell|\hat{f}| z,\ell\rangle,
\end{equation*}
where $|z,\ell\rangle$ are the Perelomov coherent states,
\begin{equation*}
 |z,\ell\rangle:=\frac{(\ah_\alpha \bz^\alpha)^\ell}{\ell!}|0\rangle.
\end{equation*}
The quantization map is given by the inverse of $\sigma_\ell$ on the set $\Sigma_\ell:=\sigma_\ell(\sEnd(\CCH_\ell))\subsetneq \CC^\infty(\CPP^n)$ of quantizable functions. Explicitly, we have
\begin{equation*}
 \sigma^{-1}_\ell\left(\frac{z_{\alpha_1}\cdots z_{\alpha_\ell}\, \bz_{\beta_1}\cdots \bz_{\beta_\ell}}{|z|^{2\ell}}\right)=\frac{1}{\ell!}\, \ah^\dagger_{\alpha_1}\cdots\ah^\dagger_{\alpha_\ell}|0\rangle\langle 0|\ah_{\beta_1}\cdots\ah_{\beta_\ell}.
\end{equation*}
Furthermore, $\sigma^{-1}_\ell(1)=\unit$ and real functions are mapped to hermitian operators in $\sEnd(\CCH_\ell)$.

Note that for $\CPP^1$, the real part of $\Sigma_\ell$ is given by the spherical harmonics with maximal angular momentum $\ell$. In general, the endomorphisms $\sEnd(\CCH_\ell)\cong \Sigma_\ell$ split into irreducible representations of of $\sSU(n+1)$ according to
\begin{equation*}
 \underbrace{\yng(6)}_{\ell}\otimes \underbrace{\overline{\yng(6)}}_{\ell} = \mathbf{1} \oplus  n 
 \left\{\phantom{\yng(2,1,1)}\right.\hspace{-1cm}\underbrace{\yng(2,1,1)}_2 \oplus\, n 
 \left\{\phantom{\yng(2,1,1)}\right.\hspace{-1cm}\underbrace{\yng(4,2,2)}_4 \oplus \cdots
\end{equation*}
or equivalently, written in terms of Dynkin labels:
\begin{equation*}
 (\ell,0,\ldots,0)\otimes\overline{(\ell,0,\ldots,0)}=(\ell,0,\ldots,0)\otimes(0,\ldots,0,\ell)
 =\oplus_{m=0}^\ell(m,0,\ldots,0,m).
\end{equation*}

The generators of $\asu(n+1)$ are represented on $\sEnd(\CCH_\ell)$ by the adjoint action of hermitian matrices $L_i$, and we introduce the quadratic Casimir operator according to
\begin{equation*}
C_2 \hat{f}:=[L_i,[L_i,\hat{f}]].
\end{equation*}
The eigenvalues of $C_2$ are positive and given on the irreducible subspace with Dynkin labels $(m,0,\ldots,0,m)$ by\footnote{Note that our conventions for $C_2$ dif\/fer from \cite{O'Connor:2007ea} by a factor of~2.} $2m(m+n)$. The degeneracy of each of these eigenspaces is given by
\begin{equation*}
 N_{n,m}^2-N_{n,m-1}^2=\frac{n(2m+n)((m+n-1)!)^2}{(m!)^2(n!)^2}.
\end{equation*}
Because of $C_2(\sigma^{-1}_\ell(f))=\sigma^{-1}_\ell(\Delta f)$, where $f\in\Sigma_\ell$ and $\Delta$ is the Laplace operator on $\CPP^n$, it is justif\/ied to identify $C_2$ with the Laplace operator on fuzzy $\CPP^n$.

The matrices $L_i$ represent the generators of $\asu(n+1)$ and thus satisfy the algebra
\begin{equation*}
 [L_i,L_j]=:\di f_{ijk}L_k,
\end{equation*}
where the $f_{ijk}$ are the structure constants of $\asu(n+1)$. We choose the $L_i$ such that
\begin{equation*}
 \tr(L_i)=0,\qquad L_i^2=c_L \unit\qquad \mbox{and}\qquad \tr(L_iL_j)=\frac{c_L N_{n,\ell}}{(n+1)^2-1}\delta_{ij}.
\end{equation*}
In the adjoint representation $R=(1,0,\ldots,0,1)$, they satisfy the Fierz identity
\begin{equation}\label{eq:FierzIdentitysun}
 L_i^{\alpha\beta}L_i^{\gamma\delta}=\delta^{\alpha\delta}\delta^{\beta\gamma}
 -\frac{1}{n+1}\delta^{\alpha\beta}\delta^{\gamma\delta},
\end{equation}
from which we conclude that
\begin{equation*}
 \tr_R(L_iL_i)=(n+1)^2-1\qquad \mbox{and}\qquad \tr_R(L_iL_j)=\delta_{ij}.
\end{equation*}
With the above relation, one readily verif\/ies the following identity for the structure constants:
\begin{equation}\label{eq:StructureConstantsIdentity}
 f_{ijk}f_{ijl}=2 (n+1)\delta_{kl}.
\end{equation}

Using the overcompleteness relation for the Perelomov coherent states,
\begin{equation*}
 \int \dif{\mu}|z,\ell\rangle\langle z,\ell|={\rm vol}(\CPP^n)\unit,
\end{equation*}
where $\dd\mu=\frac{\omega^n}{n!}$ is the Liouville measure obtained from the K{\"a}hler form $\omega$ yielding the Fubini--Study metric, one readily deduces a formula for integration: Given a function $f\in \Sigma_\ell$, the integral can be written as a trace over the quantized function $\sigma_\ell^{-1}(f)\in\sEnd(\CCH_\ell)$:
\begin{equation*}
 \int \dif{\mu}f=\frac{{\rm vol}(\CPP^n)}{N_{n,\ell}}\tr(\sigma_\ell^{-1}(f)).
\end{equation*}

\subsection[Quantum scalar field theory on CP**nF]{Quantum scalar f\/ield theory on $\boldsymbol{\CPP^n_F}$}

As we are interested in matrix models, it is convenient to switch from the label $\ell$ of our representations to the label $N_{n,\ell}$ and drop the subscript. One should, however, keep in mind that only for $\CPP^1$, there is an $\ell$ for every value of~$N$. In the following, we will represent elements of~$\sEnd(\CCH_\ell)$ by hermitian matrices $\Phi$ of dimension~$N\times N$.

In the previous section, we collected all the necessary results for writing down a scalar f\/ield theory on fuzzy $\CPP^n$. Putting everything together, we arrive at the following action functional\footnote{We implicitly reabsorbed all volume factors by a rescaling of the f\/ield $\Phi$ and the couplings.} on $\sEnd(\CCH_\ell)$:
\begin{equation}\label{eq:ActionFieldTheory}
S[\Phi]:=\tr\left(\Phi C_2\Phi+r\,\Phi^2+g\,\Phi^4\right) =
 \tr\left(\Phi[L_i,[L_i,\Phi]]+r\,\Phi^2+g\,\Phi^4\right).
\end{equation}
As we work with hermitian generators $L_i$, the quadratic Casimir operator $C_2$ has positive eigenvalues and for $r\in\FR$ and $g>0$, the action is therefore bounded from below. This, together with the f\/inite dimensionality of $\sEnd(\CCH_\ell)$, enables us to introduce the well-def\/ined functional integral
\begin{equation}\label{eq:PartitionFunction}
 \CZ:=\int \CCD \Phi~\de^{-\beta S[\Phi]}:=\int \dif{\mu_D(\Phi)}\de^{-\beta S[\Phi]},
\end{equation}
where $\dd \mu_D(\Phi)$ is the Dyson measure on the set of hermitian matrices of dimension $N\times N$.

Recall that we can diagonalize a hermitian matrix $\Phi$ according to $\Phi=\Omega\Lambda\Omega^\dagger$, where $\Omega\in\sU(N)$ and $\Lambda=\diag(\lambda_1,\ldots,\lambda_N)$ is the diagonal matrix of eigenvalues of $\Phi$. Under this decomposition, the Dyson measure splits into an eigenvalue part and an ``angular'' integration over $\sU(N)$:
\begin{equation*}
 \int \dd\mu_D(\Phi)=\int\prod_{i=1}^N\dif{\lambda_i}\Delta^2(\Lambda)\int\dd \mu_H(\Omega),
\end{equation*}
where $\dd\mu_H(\Omega)$ is the Haar measure\footnote{That is the unique measure on $\sU(N)$ which is invariant under left and right group multiplication and norma\-li\-zed according to $\int \dd\mu_H(\Omega)=1$.} and $\Delta(\Lambda)$ is the Vandermonde determinant
\begin{equation*}
 \Delta(\Lambda) := \det \big([\lambda_i^{j-1}]_{ij}\big) = \prod_{i>j} (\lambda_i-\lambda_j).
\end{equation*}

In the case of simple hermitian matrix models consisting of traces (and multitraces) over polynomials in $\Phi$, the angular integration is trivial, because $\tr(\Phi^n)=\tr(\Lambda^n)$, and reduces to a~constant volume factor. The remaining integral over the eigenvalues can then be computed by standard methods as e.g.\ the saddle point approximation or orthogonal polynomials. Here, however, the kinetic term contains the f\/ixed external matrices $L_i$ which obstruct a straightforward translation to the eigenvalue picture.

\subsection[The toy models N=n+1 on CP**n]{The toy models $\boldsymbol{N=n+1}$ on $\boldsymbol{\CPP^n}$}\label{sec:toymodel}

In the case $N=n+1$, i.e.\ when $\ell=1$ and $\sEnd(\CCH_1)$ forms the adjoint representation of $\asu(n+1)$, the kinetic term of our model \eqref{eq:ActionFieldTheory} can be evaluated explicitly by using the Fierz identity~\eqref{eq:FierzIdentitysun}. We f\/ind here that
\begin{equation}\label{eq:toymodelKineticTerm}
 \tr(\Phi C_2\Phi)= \frac{\tr(K)}{N^3-N}\big(N\tr(\Phi^2)-\tr(\Phi)\tr(\Phi)\big),
\end{equation}
where $\tr(K)$ stands for the sum over the eigenvalues of $C_2$ on $\sEnd(\CCH_1)$. Note that, as necessary, the kinetic term vanishes for $\Phi\sim\unit$. We will use this class of toy models for consistency checks of our computations below.

\section{The high-temperature expansion}\label{section3}

As it does not seem possible to compute the partition function \eqref{eq:PartitionFunction} analytically, we perform a~high-temperature expansion as suggested in~\cite{O'Connor:2007ea}. That is, we separate out the kinetic term in the functional integral and Taylor-expand its exponential, assuming $\beta$ to be small. As $\beta$ is usually inversely proportional to the temperature in statistical mechanics models, this expansion is also known as a a high-temperature expansion in the literature. For each of the terms appearing in this expansion, the integral over the angular part of the Dyson measure can be performed -- in principle straightforwardly -- using group theoretic methods. The results can be rewritten in terms of multitrace terms, and, after putting them back into the exponential of the functional integral, one ends up with a multitrace matrix model.

\subsection{Setup of the expansion}

Let us consider our model \eqref{eq:ActionFieldTheory} on fuzzy $\CPP^n_F$ with the dimension of the quantum Hilbert space~$\CCH_\ell$ being~$N$. The space of quantized functions $\sEnd(\CCH_\ell)$ is spanned by the generators\footnote{Cf.\ Appendix~\ref{appendixA} for our Lie algebra conventions.} $\tau_\mu$, $\mu=1,\ldots,N^2$ of $\au(N)$. We start by rewriting the kinetic term of the action in the following way:
\begin{equation*}
\tr(\Phi C_2 \Phi) = \tr\left(\Phi[L_i,[L_i,\Phi]]\right) = \tr(\tau_\mu[L_i,[L_i,\tau_\nu]])\,\tr(\Phi\,\tau_\mu)\,\tr(\Phi\,\tau_\nu) =: K_{\mu\nu}\Phi_\mu\Phi_\nu.
\end{equation*}
Because of $C_2\unit_N=0$, we have $K_{\mu\nu}\Phi_\mu\Phi_\nu=K_{mn}\Phi_m\Phi_n$, $m,n=1,\ldots,N^2-1$. The expansion of the kinetic term in the action now reads as
\begin{equation*}
 \de^{-\beta\tr(\Phi C_2\Phi)}=1-\beta K_{mn}\Phi_m\Phi_n+\frac{\beta^2}{2}(K_{mn}\Phi_m\Phi_n)^2-\frac{\beta^3}{6}(K_{mn}\Phi_m\Phi_n)^3+\CO(\beta^4),
\end{equation*}
and we will restrict our attention in the following to the terms up to order $\CO(\beta^3)$.

We want to perform the integral over the $\sU(N)$ part of the Dyson measure, i.e.\ to integrate out the angular degrees of freedom in $\Phi$. For this, we decompose the hermitian matrix $\Phi$ according to $\Phi=\Omega\Lambda\Omega^\dagger$, where $\Omega\in\sU(N)$ and $\Lambda=\diag(\lambda_1,\ldots,\lambda_N)$. The integrals we have to evaluate at order $\CO(\beta^k)$ are thus of the form
\begin{equation}\label{eq:HaarIntegrals}
 \CCI_k\ :=\ \int \dif{\mu_H(\Omega)}\prod_{i=1}^k K_{m_in_i}\tr(\Omega\Lambda\Omega^\dagger\tau_{m_i})\tr(\Omega\Lambda\Omega^\dagger\tau_{n_i}),
\end{equation}
where the essential part in index notation is given by
\begin{equation}\label{eq:UNintegrals}
 \int \dd\mu_H(\Omega)\, \Omega_{\alpha_1\beta_1}\cdots\Omega_{\alpha_{2k}\beta_{2k}}
 \Omega^\dagger_{\gamma_i\delta_i}\cdots\Omega^\dagger_{\gamma_{2k}\delta_{2k}}.
\end{equation}
Various algorithms have been proposed in the literature to compute integrals of the type \eqref{eq:UNintegrals}, cf.\ e.g.~\cite{Creutz:1978ub,Maekawa:1986ec,Aubert:2003ak}. The most involved integral of the form \eqref{eq:UNintegrals} which we are interested in is the one for $k=3$, which is already very dif\/f\/icult to handle by the suggested methods. Fortunately, the integrals \eqref{eq:HaarIntegrals} allow for a further simplif\/ication \cite{O'Connor:2007ea}, which is then accessible via group theoretic methods. Using $\tr(A)\tr(B)=\tr(A\otimes B)$ and $AB\otimes CD=(A\otimes C)(B\otimes D)$, we rewrite~\eqref{eq:HaarIntegrals} according to
\begin{gather*}
\CCI_k  = \int \dd\mu_H(\Omega)\,K_{m_1n_1}\cdots K_{m_kn_k}\\
\phantom{\CCI_k  =}{} \times \tr\left((\Omega\otimes\cdots\otimes\Omega)(\Lambda\otimes\cdots\otimes\Lambda)
(\Omega^\dagger\otimes\cdots\otimes\Omega^\dagger)(\tau_{m_1}\otimes\tau_{n_1}\otimes
\cdots\otimes\tau_{m_k}\otimes\tau_{n_k})\right).
\end{gather*}
The idea presented in \cite{O'Connor:2007ea} is now to use the orthogonality relation of the Haar measure \eqref{eq:OrthogonalityRelation} to evaluate these integrals. We thus have
\begin{equation*}
 \CCI_k=K_{m_1n_1}\cdots K_{m_kn_k}\sum_\rho \frac{1}{\dim(\rho)}\tr_\rho(\Lambda\otimes\cdots\otimes\Lambda)
 \tr_\rho(\tau_{m_1}\otimes\tau_{n_1}\otimes\cdots\otimes\tau_{m_k}\otimes\tau_{n_k}),
\end{equation*}
where the sum is taken over the irreducible representations contained in the tensor product of $2k$ fundamental representations of $\sSU(N)$. The traces $\tr_\rho$ are taken in the representation $\rho$, and we have
\begin{equation*}
 \tr_\rho(\Lambda\otimes\cdots\otimes\Lambda)=\chi_\rho(\Lambda),
\end{equation*}
where $\chi_\rho(\Lambda)$ denotes the character of $\Lambda$ in the representation~$\rho$. Characters of representations of $\sSU(N)$ can easily be calculated using e.g.\ the formulas in \cite{Bars:1980yy}. The remaining challenge is therefore to evaluate $\tr_\rho(\tau_{m_1}\otimes\tau_{n_1}\otimes\cdots\otimes\tau_{m_k}\otimes\tau_{n_k})$.

\subsection[The restricted traces tr rho(.)]{The restricted traces $\boldsymbol{{\rm tr}_\rho(\cdot)}$}

Consider again the generators $\tau_m$ of $\asu(N)$, and denote their matrix components by $\tau_m^{\alpha\beta}$, $\alpha,\beta=1,\ldots,N$. The full trace over the tensor products of matrices in index notation is given by
\begin{equation*}
 \tr(\tau_{m_1}\otimes\cdots\otimes\tau_{m_{2k}})=\tau_{m_1}^{\alpha_1\beta_1}\cdots
 \tau_{m_{2k}}^{\alpha_{2k}\beta_{2k}}\delta_{\alpha_1\beta_1}\cdots\delta_{\alpha_{2k}\beta_{2k}}.
\end{equation*}
To evaluate the restricted traces $\tr_\rho(\cdot)$, we need to project onto the irreducible representations which we do using projectors $\CP^{(i,j)}_{2k}$ constructed from Young symmetrizers. The technical details of the construction of these projectors are given in Appendix~\ref{appendixB}. Explicitly, we let the projec\-tor~$\CP^{(i,j)}_{2k}$ act onto the indices $\beta$ appearing in the Kronecker deltas to restrict to a representa\-tion~$\rho^{(i,j)}$:
\begin{equation*}
 \tr_{\rho^{(i,j)}}(\tau_{m_1}\otimes\cdots\otimes\tau_{m_{2k}})
 =\tau_{m_1}^{\alpha_1\beta_1}\cdots\tau_{m_{2k}}^{\alpha_{2k}\beta_{2k}}
 \CP^{(i,j)}_{2k}\delta_{\alpha_1\beta_1}\cdots\delta_{\alpha_{2k}\beta_{2k}}.
\end{equation*}
The completeness relation \eqref{eq:ProjectorCompleteness} for the projectors $\CP^{(i,j)}_{2k}$ translates into the following completeness relation for the restricted traces:
\begin{equation*}
 \sum_{i,j}\tr_{\rho^{(i,j)}}(\tau_{m_1}\otimes\cdots\otimes\tau_{m_{2k}})
 =\tr(\tau_{m_1}\otimes\cdots\otimes\tau_{m_{2k}}),
\end{equation*}
which can serve as a f\/irst consistency check of the correctness of the calculated projectors $\CP^{(i,j)}_{2k}$. A second test is to verify that each individual restricted trace indeed reduces to the character if all the $\tau_{m}$ are equal:
\begin{equation*}
 \tr_{\rho^{(i,j)}}(\Lambda\otimes\cdots\otimes\Lambda)=\chi_{\rho^{(i,j)}}(\Lambda).
\end{equation*}
Let us now compute the combined sums of the restricted traces for each type of Young tableaux and contract the $\tau_{m}$ with the $K_{mn}$ to simplify the results. That is, we compute the following expressions:
\begin{equation*}
 K_{m_1m_2}\cdots K_{m_{2k-1}m_{2k}}\tr_{\rho^{(i,j)}}(\tau_{m_1}\otimes\tau_{m_2}\otimes\cdots
 \otimes\tau_{m_{2k-1}}\otimes\tau_{m_{2k}}).
\end{equation*}
For $k=1$ and $k=2$, corresponding to the contributions at orders $\CO(\beta)$ and $\CO(\beta^2)$, these sums have already been calculated in \cite{O'Connor:2007ea}. They are
\begin{equation*}
 \begin{aligned}
   \tyng(2)&\ :\ +\frac{1}{2}\tr(K),\\
   \tyng(1,1)&\ :\ -\frac{1}{2}\tr(K),
 \end{aligned}
\end{equation*}
\begin{equation*}
\begin{aligned}
 \tyng(4) &\ :\ -\frac{20\tr(K)-(4+N)\tr(K)^2-2N\tr(K^2)}{24N},\\
\tyng(2,2) &\ :\ +\frac{(\tr K)^2+2\tr K^2}{6},\\
\tyng(3,1) &\ :\ +\frac{20\tr(K)-(4+N)\tr(K)^2-2N\tr(K^2)}{8N},\\
\tyng(2,1,1) &\ :\ -\frac{20\tr(K)+(N-4)\tr(K)^2+2N\tr(K^2)}{8N},\\
\tyng(1,1,1,1) &\ :\ +\frac{20\tr(K)+(N-4)\tr(K)^2+2N\tr(K^2)}{24N}.
 \end{aligned}
\end{equation*}
The lengthy result for $k=3$ is given in Appendix~\ref{appendixC}.

In calculating these results, we used many identities which we will brief\/ly comment on now: First of all, using the Fierz identity for the generators of $\au(N)$ as well as the relations for the~$L_i$, we compute that for arbitrary $A,B\in\au(N)$,
\begin{gather}
 K_{\mu\nu}\tr(\tau_\mu A)\tr(\tau_\nu B) = 2 c_L \tr(AB)-2\tr(L_iAL_iB),\nonumber\\
 K_{\mu\nu}\tr(\tau_\mu A\tau_\nu B) = 2 c_L \tr(A)\tr(B)-2\tr(L_iA)\tr(L_iB).\label{eq:K-resolutions}
\end{gather}
Applying these relations to $\tr(K):=K_{\mu\nu}\tr(\tau_\mu\tau_\nu)$ yields
\begin{equation*}
 c_L=\frac{\tr(K)}{2N^2}\qquad \mbox{and}\qquad  d_g:=(n+1)^2-1=\frac{\tr(K)^2}{N^2\tr(K^2)-\tr(K)^2}.
\end{equation*}
The identities \eqref{eq:K-resolutions} allow us to successively rewrite expressions involving $K_{\mu\nu}$ in terms of traces over products of the $L_i$, which in turn can be reduced using $L_i^2=c_L\unit_N$ and the identity for the structure constants~\eqref{eq:StructureConstantsIdentity}. Some useful intermediate results are collected in Appendix~\ref{appendixC}.

\subsection{The multitrace matrix model}

Combining the reduced traces with the characters in the various representations, we arrive at the following expressions for the $\CCI_k$:
\begin{gather*}
 \CCI_1=\frac{\tr(K)}{N^2-1}\tr(\Lambda^2)-\frac{\tr(K)}{N^3-N}\tr(\Lambda)^2,\\
 \CCI_2= \frac{10 \tr(K) \left(-2 \left(1+N^2\right)+\tr(K)\right)+4 \left(3-2 N^2\right) \tr\left(K^2\right)}{N \left(-36+N^2 \left(-7+N^2\right)^2\right)}\tr(\Lambda^4)\\
\phantom{\CCI_2=}{}+\frac{40 \left(2+2 N^2-\tr(K)\right) \tr(K)+16 \left(-3+2 N^2\right) \tr\left(K^2\right)}{N^2 \left(-36+N^2 \left(-7+N^2\right)^2\right)}\tr(\Lambda^3)\tr(\Lambda)\\
\phantom{\CCI_2=}{}+ \frac{20 \left(-3\!+\!2 N^2\right) \tr(K)\!+\!\left(30\!-\!14 N^2\!+\!N^4\right) \tr(K)^2\!+\!2 \left(18\!-\!6 N^2\!+\!N^4\right) \tr\left(K^2\right)}{N^2 \left(-36+N^2 \left(-7+N^2\right)^2\right)}\tr(\Lambda^2)^2\\
\phantom{\CCI_2=}{}-\frac{2 \left(100 \tr(K)+\left(-14+N^2\right) \tr(K)^2+2 \left(6+N^2\right) \tr\left(K^2\right)\right)}{N \left(-36+N^2 \left(-7+N^2\right)^2\right)}\tr(\Lambda^2)\tr(\Lambda)^2\\
\phantom{\CCI_2=}{}+\frac{100 \tr(K)+\left(-14+N^2\right) \tr(K)^2+2 \left(6+N^2\right) \tr\left(K^2\right)}{N^2 \left(-36+N^2 \left(-7+N^2\right)^2\right)}\tr(\Lambda)^4.
\end{gather*}
The result for $\CCI_3$ is lengthy and because it can be easily calculated from the list of restricted traces given in Appendix~\ref{appendixC}, we refrain from presenting it here. Note that all the above integrals pass the f\/irst consistency check: We have $\CCI_k=0$ if $\Lambda\sim\unit_N$.

To rephrase the perturbative expansion in terms of an ef\/fective action, we re-exponentiate the terms. That is, we write
\begin{equation*}
 \de^{-\beta(S_1+S_2+S_3)}=1-\beta\CCI_1+\frac{\beta^2}{2}\CCI_2-\frac{\beta^3}{6}\CCI_3+\CO(\beta^4),
\end{equation*}
where we demand that the $S_i$ are polynomials in the eigenvalues of order $2i$. As the same holds by def\/inition for the $\CCI_i$, we can match both sides order by order and arrive at
\begin{equation*}
 S_1=\CCI_1, \qquad S_2=\frac{\beta}{2}(\CCI_1^2-\CCI_2), \qquad S_3=\frac{\beta^2}{6}(2\CCI_1^3-3\CCI_1\CCI_2+\CCI_3).
\end{equation*}
We can now perform the second consistency check of our result and compare the re-exponen\-tiated action with the toy model $N=n+1$ from Section~\ref{sec:toymodel}. In the representation $N=n+1$, we have
\begin{gather*}
 \tr(K)=2n(1+n)(2+n),\qquad\!\!\! \tr(K^2)=4n(1+n)^2(2+n),\qquad\!\!\! \tr(K^3)=8n(1+n)^3(2+n).\!
\end{gather*}
Plugging this into the expressions for $S_1,S_2$ and $S_3$, we f\/ind that $S_1$ indeed reduces to the kinetic term of the toy model~\eqref{eq:toymodelKineticTerm}. Since the terms $S_2$ and $S_3$ vanish as required, our results pass this consistency check as well.

By re-inserting the integration over the Haar measure, we can return from the eigenvalues to the full hermitian matrices $\Phi$. We thus obtain the multitrace matrix model with action $S=S_1+S_2+S_3$ with
\begin{equation*}
  S_1= \frac{\tr(K)}{N^2-1}\tr\big(\Phi^2\big)-\frac{\tr(K)}{N^3-N}\tr(\Phi)^2,
\end{equation*}
etc. The involved expressions for $S_2$ and $S_3$ are again lengthy but easily calculated from the results given above.

Altogether, we obtained a multitrace matrix model whose partition function approximates the partition function of fuzzy scalar f\/ield theory on complex projective space up to order $\CO(\beta^3)$. This approximation should be valid in particular for large values of the couplings $r$ and $g$.

\section[Large N solutions of the model]{Large $\boldsymbol{N}$ solutions of the model}\label{section4}

The partition function of the multitrace matrix model we obtained in the previous section can now be evaluated analytically for f\/inite $N$ using the methods of orthogonal polynomials. As we are mainly interested in the phase diagram, we consider instead the large $N$ limit and use the saddle point method to determine the partition function here. We try to be self-contained and present the involved steps in detail.

\subsection[The large N limit]{The large $\boldsymbol{N}$ limit}

The phase diagram determined numerically in \cite{GarciaFlores:2005xc} is invariant under the multiscaling limit
where $N\rightarrow\infty$ and $N^2\beta g$ as well as $N^{3/2}\beta r$ are kept f\/ixed. This justif\/ies to solve our model in the large $N$ limit to compare it with the phase diagram. In this limit, the discrete set of eigenvalues goes over into a continuous function: We rescale $\lambda_i\rightarrow \lambda(i/N)=:\lambda(x)$, with $0< x \leq 1$. The traces turn correspondingly into integrals: $\tr(\Phi^j)=\sum_i\lambda_i^j\rightarrow N\int_0^1\dd x\,\lambda(x)^j$.

The formulas for general $n$ turn out to be very lengthy and dif\/f\/icult to handle. Therefore we will restrict our attention in the following to the three projective spaces $\CPP^1$, $\CPP^2$ and~$\CPP^3$. The f\/irst case $n=1$ is interesting, as we would like to compare the resulting phase diagram to the one numerically obtained in~\cite{GarciaFlores:2005xc}. The second case $n=2$ is a well-def\/ined, four-dimensional quantum f\/ield theory with quartic potential. The third case $n=3$ is interesting as the corresponding scalar f\/ield theory on $\FR^6$ is not renormalizable.

The eigenvalues of the quadratic Casimir, the degeneracy of the corresponding eigenspaces and the dimension of the representation $\overline{(\ell,0,\ldots,0)}\otimes(\ell,0,\ldots,0)$ for the cases $n=1,2,3$ are listed in the following table:
\begin{center}
\begin{tabular}{r|c|c|c}
& $\CPP^1$ & $\CPP^2$ & $\CPP^3$ \\
\hline
 eigenvalues of $C_2$ & $2\ell(\ell+1)$ & $2\ell(\ell+2)$ & $2\ell(\ell+3)$\tsep{2pt}\\
 degeneracy of eigenspaces & $1+2\ell$ & $(1+\ell)^3$ & $\tfrac{1}{12}(1+\ell)^2(2+\ell)^2(3+2\ell)$\\
 $N_{n,\ell}$ & $\ell+1$ & $\tfrac{1}{2}(\ell+1)(\ell+2)$ & $\tfrac{1}{6}(\ell+1)(\ell+2)(\ell+3)$\\
\end{tabular}
\end{center}
As the function $N_{n,\ell}$ is only surjective for $n=1$, and thus we cannot f\/ind an $\ell$ for every value of $N$, we will rewrite the multitrace matrix model in terms of~$\ell$.

From the table above, we easily evaluate the various traces over $K$ appearing in the action of the multitrace matrix model. We have for $\CPP^1$:
\begin{gather*}
 \tr(K)=\ell(1+\ell)^2 (2+\ell),\qquad \tr(K^2)=\tfrac{4}{3}\ell^2(1+\ell)^2(2+\ell)^2,\\
 \tr(K^3) =\tfrac{2}{3}\ell^2(1+\ell)^2(2+\ell)^2(3\ell^2+6\ell-1),
\end{gather*}
for $\CPP^2$:
\begin{gather*}
 \tr(K)=\tfrac{1}{3}\ell(1+\ell)^2 (2+\ell)^2(3+\ell),\qquad
 \tr(K^2)=\tfrac{1}{2}\ell^2(1+\ell)^2(2+\ell)^2(3+\ell)^2,\\
\tr(K^3) =\tfrac{1}{5}\ell^2(1+\ell)^2(2+\ell)^2(3+\ell^2)(4\ell^2+12\ell-1),
\end{gather*}
and for $\CPP^3$:
\begin{gather*}
 \tr(K) =\tfrac{1}{24}\ell(1+\ell)^2(2+\ell)^2(3+\ell)^2(4+\ell),\\
\tr(K^2) =\tfrac{1}{15}\ell^2(1+\ell)^2(2+\ell)^2(3+\ell)^2(4+\ell)^2,\\
\tr(K^3) =\tfrac{1}{45}\ell^2(1+\ell)^2(2+\ell)^2(3+\ell)^2(4+\ell)^2(5\ell^2+20\ell-1).
\end{gather*}

In the limit $\ell\rightarrow\infty$, the expressions for the various matrix models simplify. Switching to the eigenvalue description and using the moments
\begin{equation*}
 c_n:=\int \dd x\, \lambda^n(x),
\end{equation*}
we can write them down explicitly. On the three fuzzy $\CPP^n$s, we have the models
\begin{gather*}
 \beta S^{(n=1)}=
 \beta\ell^3(c_2-c_1^2)-\beta^2\frac{\ell^4}{3}
 \left(c_1^2-c_2\right)^2-\beta^3\frac{4\ell^5}{27}\left(2c_1^3-3c_1c_2+c_3\right)^2\\
 \phantom{\beta S^{(n=1)}=}{}
 +\beta \ell r c_2+\beta \ell g c_4-\ell^2\int\dd x\,\dd y\,\log|\lambda(x)-\lambda(y)|,\\
\beta S^{(n=2)}= \beta\frac{2\ell^4}{3}(c_2-c_1^2)-\beta^2\frac{2\ell^4}{9}\left(c_1^2-c_2\right)^2\\
\phantom{\beta S^{(n=2)}=}{} - \beta^3\frac{8\ell^4}{405}\left(12c_1^6-36c_1^4c_2+21c_1^2c_2^2+8c_2^3+10c_1(2c_1^2-3c_2)c_3+5c_3^2\right)\\
\phantom{\beta S^{(n=2)}=}{} +\beta\frac{\ell^2}{2}r c_2+\beta\frac{\ell^2}{2}g c_4-\frac{\ell^4}{4}\int\dd x\,\dd y\,\log|\lambda(x)-\lambda(y)|,\\
\beta S^{(n=3)}= \beta\frac{\ell^5}{4}\left(c_2-c_1^2\right)-\beta^2\frac{3\ell^4}{20}\left(c_1^2-c_2\right)^2\\
\phantom{\beta S^{(n=3)}=}{}
- \beta^3\frac{\ell^3}{25}\left(2c_1^6-6c_1^4c_2-3c_1^2c_2^2+10c_2^3+6c_1(2c_1^2-3c_2)c_3+3c_3^2\right)\\
\phantom{\beta S^{(n=3)}=}{}
+\beta\frac{\ell^3}{6} r c_2+\beta\frac{\ell^3}{6} g c_4-\frac{\ell^6}{36}\int\dd x\,\dd y\,\log|\lambda(x)-\lambda(y)|,
\end{gather*}
where the repulsive log-term arises as usual from exponentiating the Vandermonde determinant. The $\ell$-dependence of the log-term is due to the factor $N_{n,\ell}^2$, which in turn originated from rewriting the double sum as a double integral. In the above expressions, subleading terms in $\ell$ have been suppressed in each summand.

As a next step, we have to f\/ind the appropriate multi-scaling behavior of the constants~$\beta$,~$r$,~$g$ and the continuous eigenvalues $\lambda(x)$. The coef\/f\/icients of the log-terms determines the desired scaling behavior of the total action. We f\/ix the remaining scalings by demanding that the whole action scales homogeneously and that $\beta g$ scales with $N^2$, as for an ordinary hermitian matrix model. We thus f\/ind the following rescalings:
\begin{equation*}
 \begin{tabular}{c|llll}
   $\CPP^1$ & $\beta\rightarrow \ell^{-\frac{1}{2}}\beta$, & $\lambda(x)\rightarrow \ell^{-\frac{1}{4}}\lambda(x)$, & $r\rightarrow \ell^{2}r$, & $g\rightarrow \ell^{\frac{5}{2}}g$\phantom{\Big(} \\
\hline
   $\CPP^2$ & $\beta\rightarrow \beta$, & $\lambda(x)\rightarrow \lambda(x)$, & $r\rightarrow \ell^2r$, & $g\rightarrow \ell^2g$\phantom{\Big(} \\
\hline
   $\CPP^3$ & $\beta\rightarrow \ell^{\frac{1}{2}}\beta$, & $\lambda(x)\rightarrow \ell^{\frac{1}{4}}\lambda(x)$, & $r\rightarrow \ell^{2}r$, & $g\rightarrow \ell^{\frac{3}{2}}g$\phantom{\Big(} \\
 \end{tabular}
\end{equation*}
Note that the scalings for $\CPP^1$ indeed agree with the ones numerically determined in \cite{GarciaFlores:2005xc} as well as the ones calculated in \cite{O'Connor:2007ea}.

As a f\/inal simplif\/ication, we note that our theory is invariant under $\Phi\rightarrow -\Phi$, as the potential is even. We expect the eigenvalues to respect this symmetry\footnote{As we will see later, this assumption is not correct in a small part of the conf\/iguration space. At this point, it serves as a very useful approximation to keep the terms in the action manageable.}, and therefore we put all the odd moments $c_{2n+1}$, $n\in\NN$ to zero. Moreover, we replace the integral over $x$ by an integral over the eigenvalue density $\rho(\lambda):=\frac{\dd x}{\dd \lambda}$. We thus eventually arrive at the following three models, which we wish to solve:
\begin{gather*}
 \beta S^{(n=1)}= \beta \left(1-\frac{\beta}{3}c_2+r\right)c_2+\beta g c_4-\int\dd \lambda\,\dd \mu\,\rho(\lambda)\log|\lambda-\mu|\rho(\mu),\\
\beta S^{(n=2)}= \beta\left(\frac{8}{3}-\frac{8\beta}{9}c_2-\frac{256\beta^2}{405}c_2^2+2r\right)c_2+2\beta g c_4-\int\dd \lambda\,\dd \mu\,\rho(\lambda)\log|\lambda-\mu|\rho(\mu),\\
\beta S^{(n=3)}= \beta\left(9-\frac{27\beta}{5}c_2-\frac{72\beta^2}{5}c_2^2+6r\right)c_2+6\beta g c_4-\int\dd \lambda\,\dd \mu\,\rho(\lambda)\log|\lambda-\mu|\,\rho(\mu).
\end{gather*}

\subsection{Solving the models}

We will now calculate the partition functions of our models using the saddle point method: In the large $\ell$ limit, the path integral localizes on classical solutions, or saddle points, of the action\footnote{Note that we had to switch to the eigenvalue formulation of the matrix models f\/irst, as the zero modes corresponding to the angular degrees of freedom contained in $\Phi$ would have rendered the approximation invalid.}. These solutions, which are valid only for a restricted range of the coupling constants, can be easily obtained using standard methods in random matrix theory. We start from the action\footnote{We have included a Lagrange multiplier $\xi$ to f\/ix the normalization of the eigenvalue density.}
\begin{equation*}
 S[\rho(\lambda)]=\int_\CI\dd\lambda\,\rho(\lambda)V(\lambda)-\int_{\CI\times \CI}\dd\lambda\,\dd\mu\,\rho(\lambda)\log|\lambda-\mu|\,\rho(\mu)+\xi\left(\int_\CI\dd\lambda\,\rho(\lambda)-1\right),
\end{equation*}
where $\CI$ is the union of open intervals on the real line over which $\rho(\lambda)$ has support. The saddle point equation is obtained by varying the above equation with respect to $\rho(\lambda)$:
\begin{equation}\label{eq:presaddlepoint}
 V(\lambda)-2\int_\CI\dd\mu\,\rho(\mu)\,\log|\lambda-\mu|+\xi=0.
\end{equation}
Note that our potentials satisfy $V(\lambda)=0$ at $\lambda=0$ and we can therefore determine the Lagrange multiplier $\xi$ by solving the saddle point equation at this special point if $0\in\CI$:
\begin{equation*}
 \xi_{0\in\CI}=2\int_\CI\dd\mu\,\rho(\mu)\log|\mu|;
\end{equation*}
otherwise, one has to choose a dif\/ferent value of $\lambda$ to obtain $\xi$. We def\/ine the free energy~$F$ as $F:=-\log(\CZ)$, where $\CZ$ is the partition function of our model. In the saddle point approximation, this reduces to $F=\beta S[\rho(\lambda)]$, which we can evaluate using \eqref{eq:presaddlepoint}:
\begin{equation*}
 F=\tfrac{1}{2}\int_\CI \dd \lambda\, \rho(\lambda) V(\lambda)-\tfrac{1}{2}\xi.
\end{equation*}

To f\/ind the eigenvalue density $\rho(\lambda)$, it is convenient to replace \eqref{eq:presaddlepoint} with its  derivative\footnote{When doing this, one obviously has to vary each moment: $\delta c_2^2=2c_2\delta c_2$, etc.} with respect to $\lambda$:
\begin{equation}\label{saddlepoint}
 V'(\lambda) =
2\int_\CI \hspace{-0.44cm}-~\dd \mu\,\frac{\rho(\mu)}{\lambda-\mu}.
\end{equation}
This is a singular integral equation, and its general solution can be found e.g.\ in~\cite{9001607004}, see also~\cite{DiFrancesco:1993nw}. First of all, one introduces the {\em resolvent} $W(\lambda)$, which is an analytic function on $\FC\backslash \CI$, def\/ined according to
\begin{equation*}
W(\lambda):=\int \dd \mu\, \frac{\rho(\mu)}{\lambda-\mu}.
\end{equation*}
Note that for large $\lambda$, we have $W(\lambda)\sim \frac{1}{\lambda}$. The resolvent is related to the eigenvalue densi\-ty~$\rho(\lambda)$ and the Cauchy principal value appearing in the equation of motion through the Plemelj formula, and we arrive at
\begin{gather*}
  \rho(\lambda)=-\frac{1}{2\pi\di}(W(\lambda+\di\eps)-W(\lambda-\di\eps)),\\
  V'(\lambda)=W(\lambda+\di\eps)+W(\lambda-\di\eps).
 \end{gather*}
The f\/irst equation determines $\rho(\lambda)$ in terms of the resolvent, and the second equation is a much simpler equation than \eqref{saddlepoint}, which f\/ixes the resolvent\footnote{Strictly speaking, it f\/ixes the resolvent only up to regular terms, which, however, are absent as can be seen from the large $\lambda$ behavior $W(\lambda)\sim\frac{1}{\lambda}$.} and thus the eigenvalue density. One can show that the resolvent satisf\/ies the Schwinger--Dyson equation
\begin{equation*}
 W^2(\lambda)-V'(\lambda)W(\lambda)+\tfrac{1}{4}R(\lambda)=0,
\end{equation*}
where
\begin{equation*}
 R(\lambda)=4\int \dd \mu\, \rho(\mu)\frac{V'(\lambda)-V'(\mu)}{\mu-\lambda}
\end{equation*}
is a polynomial of degree $d-2$. The solution to the above equation reads as
\begin{equation*}
 W(\lambda)=\tfrac{1}{2}(V'(\lambda)\pm \underbrace{\sqrt{V'{}^2(\lambda)-R(\lambda)}}_{:=\omega(\lambda)}),
\end{equation*}
where $\omega(\lambda)$ describes the part of $W(\lambda)$ containing the branch cuts.

Explicit solutions are now obtained by making assumptions about the support $\CI$ of the eigenvalue density $\rho(\lambda)$.
The simplest assumption is that $\CI$ consists of a single interval. This is expected if the potential either consists of one deep well or if the eigenvalue f\/illing is such that all the local minima of the potential are more than f\/illed up. In this case, the resolvent has to have a branch cut over $\CI:=(\delta_1,\delta_2)$ and the corresponding solution is therefore known as a~{\em single-cut solution}. The resolvent's singular part has to contain exactly two roots, cf.\ e.g.~\cite{Brezin:1977sv}:
\begin{equation*}
 \omega^2(\lambda)=M^2(\lambda)(\lambda-\delta_1)(\lambda-\delta_2)=V'{}^2(\lambda)-R(\lambda).
\end{equation*}
One can now make a general ansatz for the polynomials $M(\lambda)$ and $R(\lambda)$. Together with the self-consistency condition that all the moments $c_n$ satisfy their def\/ining relation
\begin{equation*}
 c_n := \int \dd \lambda\, \rho(\lambda) \lambda^n,
\end{equation*}
we can solve for all unknowns and determine $\rho(\lambda)$. Note that the normalization condition on the eigenvalue density $c_0=1$ is equivalent to the less involved condition that the asymptotic behavior of the resolvent is $W(\lambda)=\frac{1}{\lambda}+\CO(\frac{1}{\lambda^2})$.

When having a double well potential, we also expect solutions where $\CI$ is given by the union of two disjoint intervals $\CI=(\delta_1,\delta_2)\cup(\eps_1,\eps_2)$. Correspondingly, the singular part of the potential contains four roots and we make the ansatz:
\begin{equation*}
 \omega^2(\lambda)=M^2(\lambda)(\lambda-\delta_1)(\lambda-\delta_2)
 (\lambda-\eps_1)(\lambda-\eps_2)=V'{}^2(\lambda)-R(\lambda).
\end{equation*}
This solution is known as a {\em double-cut solution}.

It is important to stress that in general, all solutions will be valid only on a subset of the full parameter space of the model under consideration. This subset is characterized by the condition $\rho(\lambda)\geq 0$ (and therefore $c_{2n}\geq 0$) as well as the condition that $\CI$ is of the assumed form. It is called the {\em existence domain} of a solution, and its boundary in parameter space can correspond to a phase transition.

In the following, we will present all the solutions for the various models together with their existence domains. For $\CPP^1$, we also give the explicit expressions for the free energies. We will consider three kinds of solutions: the symmetric single-cut solution, the symmetric double-cut solution and the asymmetric single-cut solution. In the latter case, we should strictly speaking include all the odd moments $c_{2n-1}$, $n\in\NN$, which we dropped in our actions. However, the full action would be very dif\/f\/icult to handle analytically, and we hope to make at least qualitative statements with our truncation.

The solutions for closely related models, in which $r$ is kept f\/ixed, the coef\/f\/icient of $c_2^2$ is a~parameter and the coef\/f\/icient of $c_2^3$ vanishes, have been computed in \cite{Das:1989fq} for the symmetric single-cut type and \cite{Shishanin:2007zz} for the two other types.

\subsection[Solutions of the model on CP**1]{Solutions of the model on $\boldsymbol{\CPP^1}$}

For the symmetric single-cut case, we assume that the eigenvalue density $\rho(\lambda)$ has support on the interval $\CI=(-d,+d)$. The solution we obtain from the procedure described above together with the conditions $W(z)\sim \frac{1}{z}$ and $\int_\CI \dd\lambda\,\lambda^2\,\rho(\lambda)=c_2$ reads as
\begin{gather}
 c_2=\frac{3(2d^2r\beta+3d^4g\beta+2d^2\beta-4)}{4d\beta^2} ,\qquad \rho(\lambda)= \frac{\sqrt{d^2-\lambda^2}(4-d^2g\beta(d^2-4\lambda^2)}{2d^2\pi} ,\nonumber\\
48+d^2\beta(d^6g\beta^2+4d^2(\beta-9g)- 24(1+r))=0.\label{eq:solCP11CS}
\end{gather}
The free energy of this solution is given by
\begin{equation*}
F=\frac{1}{64} \left(40+d^2 \beta  \left(-12 d^2 g+4 (1+r)+d^4 g (1+r) \beta \right)-64 \log\left(\frac{d}{2}\right)\right).
\end{equation*}
The solution exists, if both $\rho(\lambda)$ and $c_2$ are nowhere negative. The condition $\rho(\lambda)\geq 0$ amounts to
\begin{equation*}
 r>-1+\frac{2(\beta-3g)}{3\sqrt{\beta g}},
\end{equation*}
while $c_2\geq 0$ is always satisf\/ied in the existence domain of the solution~\eqref{eq:solCP11CS}.

Next, we assume a symmetric double-cut support for $\rho(\lambda)$ on $\CI=(-\sqrt{s+d},-\sqrt{s-d})\cup(\sqrt{s-d},\sqrt{s+d})$. The solution here reads as
\begin{gather*}
  \rho(\lambda)=\frac{2}{\pi}g\beta\lambda\sqrt{(s+d-\lambda^2)(\lambda^2-s+d)} ,\qquad c_2=s,\qquad d=\frac{1}{\sqrt{\beta g}},\qquad s=\frac{3(1+r)}{2(\beta-3g)}.
\end{gather*}
To evaluate the free energy, we compute $\xi$ at $\lambda=\sqrt{s}$ and use the relation
\begin{equation*}
 F=\tfrac{1}{2}\left(\int_\CI\dif{\lambda}\rho(\lambda)
 \left(V(\lambda)-\log(\lambda-\sqrt{s})-\log(\lambda+\sqrt{s})\right)\right)
 +\tfrac{1}{4}V(\sqrt{s})+\tfrac{1}{4}V(-\sqrt{s}).
\end{equation*}
The result is
\begin{equation*}
 F=\frac{9 g-3 \left(3+4 r+2 r^2\right) \beta +(6 g-2 \beta ) \log(4 g \beta )}{24 g-8 \beta }.
\end{equation*}

The symmetric double-cut solution exists, if $s=c_2>0$, i.e.\ if $r<-1$ and $3g>\beta$ or $r>-1$ and $3g<\beta$ and if $s>d$. The latter condition yields
\begin{equation*}
 r<-1+\frac{2(\beta-3g)}{3\sqrt{\beta g}},
\end{equation*}
and we see that the boundary of the existence domain of the symmetric double-cut solution matches that of the symmetric single-cut solution. We therefore expect a phase transition at this boundary. At the point $(r_0,g_0)=(-1,\frac{\beta}{3})$, something interesting happens: Here, the equation for $s$ becomes trivially satisf\/ied and $s$ is unconstrained. The two cuts can thus be arbitrarily far apart. At this point, the action reduces to
\begin{equation*}
 S_{(r_0,g_0)}=-\frac{\beta}{3}c_2^2+\frac{\beta}{3}c_4,
\end{equation*}
and we have a competition of the single-trace and the multitrace potential term.

The asymmetric single-cut solution with support on $\CI=(s-d,s+d)$ with $s\neq 0$ is given by
\begin{gather*}
  \rho(\lambda)=\frac{2 g \beta  \sqrt{d^2-(s-\lambda )^2} \left(\lambda  (s+\lambda )-d^2\right)}{\pi } , \qquad c_2=\frac{3(1+3d^2g+r+2gs^2)}{2\beta}, \\
s=\sqrt{\frac{4+3d^4g\beta}{8d^2g\beta}} ,\qquad 4 \beta +g \left(-12-12 d^2 (1+r) \beta +5 d^8 g \beta ^3+d^4 \beta  (-45 g+11 \beta )\right)=0.
 \end{gather*}
To evaluate the free energy, we determine the Lagrange multiplier $\xi$ at $\lambda=s$. Our def\/initions then yield
\begin{equation*}
 F=\frac{8 (1+r)+d^2 g \left(-24+d^2 \beta  \left(-3 d^2 g+14 (1+r)+5 d^4 g (1+r) \beta \right)\right)+32 d^2 g \log\left(\frac{d}{2}\right)}{32 d^2 g}.
\end{equation*}
The asymmetric single-cut solution exits if $\rho\geq 0$ and if $d$ is real and positive. The f\/irst condition implies
\begin{equation*}
r<\frac{-135 \sqrt{15} g-135 \sqrt{g} \sqrt{\beta }+41 \sqrt{15} \beta }{135 \sqrt{g} \sqrt{\beta }},
\end{equation*}
and $c_2$ and $s$ are automatically positive. The second condition amounts to $g>\frac{\beta}{3}$.

\subsection[Phase structure on CP**1]{Phase structure on $\boldsymbol{\CPP^1}$}

\begin{figure}[h]
\centerline{\includegraphics{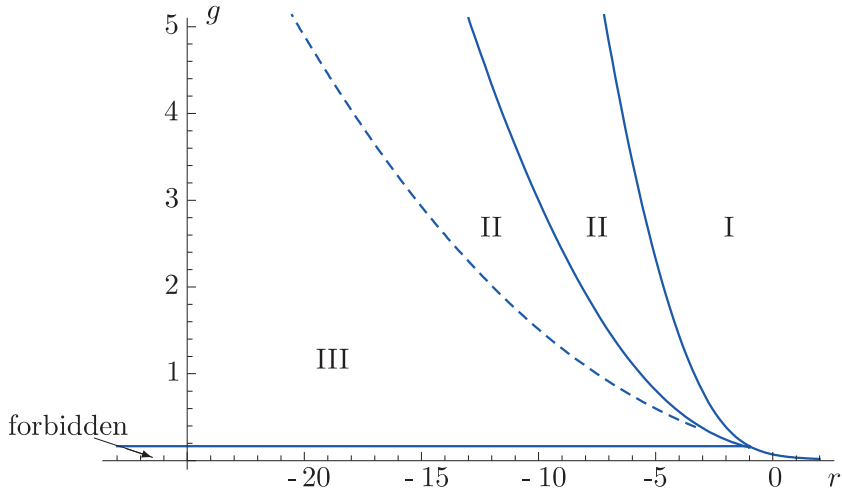}}

\caption{The phase diagram on $\CPP^1$ for $\beta=\frac{1}{2}$. Dashed lines describe phase boundaries, solid lines boundaries of existence domains. See the text for more details.}\label{fig1}
\end{figure}

If existence domains do not overlap, we expect a phase transition at the boundary. If, however, two solutions exist for the same parameters, then the solution with the lowest free energy will be adopted. The resulting phase diagram is depicted in Fig.~\ref{fig1}.

Here, the symmetric single-cut, the double-cut and the asymmetric single-cut solutions are labelled as I, II and III. The boundary of the existence domain between I and II describes the usual second order phase transition of the hermitian matrix model. The existence domain of III is fully contained in the existence domain of II. There is indeed a region of the parameter space, where the asymmetric f\/illing yields a lower value for the free energy than the symmetric f\/illing. This is particularly interesting, as it was very dif\/f\/icult to extract the II--III phase transition from the numerical data in~\cite{GarciaFlores:2005xc}. We can thus conf\/irm the numerical f\/indings. Furthermore, there is the forbidden region in the parameter space for $g<\frac{\beta}{3}$. We expect that higher-order corrections in $\beta$ would deform this boundary.

Altogether, we have obtained the general features of the phase diagram found in \cite{GarciaFlores:2005xc}: We have three distinct phases, which come together at the point $(r_0,g_0)=(-1,1/6)$, which corresponds to $(b,c)=(-0.5,1/12)$ in the conventions of \cite{GarciaFlores:2005xc}. This compares to numerically found values of $(b,c)=(-0.8\pm 0.08,0.15\pm 0.05)$. The discrepancy is due to the fact that the triple point is in a region of the parameter space, where the kinetic term is not small compared to the potential terms.

The ef\/fects due to the asymmetric single-cut region have to be considered only qualitatively, because we have dropped all the odd moments from the action using symmetry arguments. This was done to keep the solutions under analytical control. The critical line found in \cite{GarciaFlores:2005xc} corresponds here to the dashed curve. Including the odd momenta would presumably straightened this curve.

The discrepancies compared to \cite{O'Connor:2007ea} arise from the fact that there, a contribution to the action, labelled $K\urcorner K$, was neglected in the large $N$ limit while we included it here. This yielded a~dif\/ferent model with the opposite sign of the $c_2^2$ term.

\subsection[Solutions of the model on CP**2]{Solutions of the model on $\boldsymbol{\CPP^2}$}

Let us now be brief in repeating the analysis for $\CPP^2$: The single-cut solution with support on $\CI=(-d,d)$ is given by
\begin{gather*}
 \rho(\lambda)=\frac{2 \beta  \sqrt{d^2-\lambda ^2} \left(-120 c_2 \beta -128 c_2^2 \beta ^2+45 \left(4+3 d^2 g+3 r+6 g \lambda ^2\right)\right)}{135 \pi } ,\\c_2=\tfrac{1}{8} (2d^2+d^6g\beta) ,\qquad
-270+d^2 \beta  (405 d^2 g+270 r-8 (-45+2 c_2 \beta  (15+16 c_2 \beta )))=0 ,
 \end{gather*}
and the boundary of its existence domain is
\begin{equation*}
 r>-\frac{4}{3}-\sqrt{\frac{2g}{\beta }}+\frac{64 \beta +60 \sqrt{2g \beta }}{135 g} .
\end{equation*}

The double-cut solution with support $\CI=(-\sqrt{s+d},-\sqrt{s-d})\cup(\sqrt{s-d},\sqrt{s+d})$  reads as:
\begin{gather*}
  \rho(\lambda)=\frac{2 \lambda  \sqrt{g \beta  \big(2-4 g \beta  \left(s-\lambda ^2\right)^2\big)}}{\pi } ,\qquad c_2 =s ,\\
  d=\frac{1}{\sqrt{2\beta g}} ,\qquad -128 s^2+45 (4+3 r+6 g s) \beta ^2-120  s \beta ^3=0 ,
 \end{gather*}
and this solution is admissible, if $s$ is real and $s>d$. This yields the following bounds:
\begin{equation*}
\frac{-405 g^2+360 g \beta -592 \beta ^2}{384 \beta ^2}<r<-\frac{4}{3}-\sqrt{\frac{2g}{\beta }}+\frac{64 \beta +60 \sqrt{2g \beta }}{135 g}.
\end{equation*}
Note that the left and the right bound touch at a single point in the $r$-$g$-plane.

The asymmetric single-cut solution has support $\CI=(s-d,s+d)$ and is given by
\begin{gather*}
  \rho(\lambda)=\frac{4 g \beta  \sqrt{d^2-(s-\lambda )^2} \left(\lambda  (s+\lambda )-d^2\right)}{\pi },\qquad
  c_2=\frac{1}{4}d^2g(16s^4+10d^2s^2-d^4),\\
s=\sqrt{\frac{2+3d^4g\beta}{8d^2g\beta}},\qquad 45 \left(4+9 d^2 g+3 r+6 g s^2\right)-120 c_2 \beta -128 c_2 \beta ^2=0.
 \end{gather*}
This solution is valid for
\begin{equation*}
r>-\frac{4}{3}-\frac{\sqrt{15 g}}{\sqrt{2 \beta }}+\frac{82 \sqrt{2\beta }}{27 \sqrt{15g}}+\frac{26896 \beta }{18225 g}.
\end{equation*}
The moment $c_2$ and the center of the cut $s$ are automatically positive. Contrary to the case of~$\CPP^1$, there is no upper bound on $r$ for this solution.

The expressions for the free energies on~$\CPP^2$ are not presented as they are lengthy but can be calculated quite straightforwardly.

\begin{figure}[h]
\centerline{\includegraphics{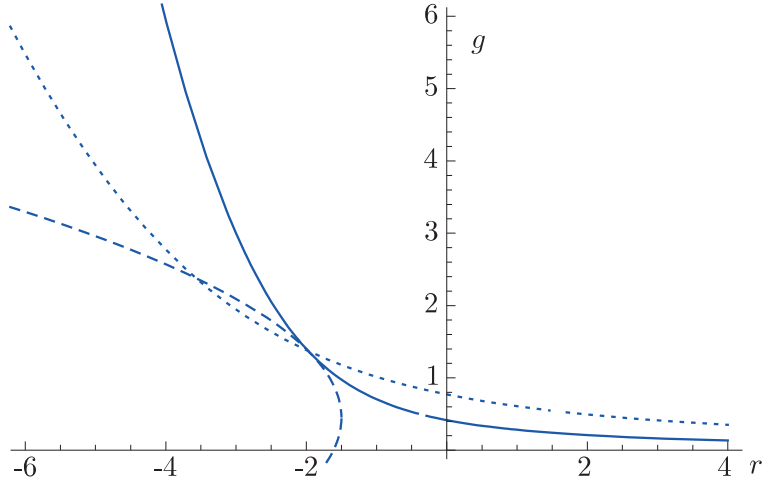}}
\caption{The boundaries of the various existence domains on $\CPP^2$ for $\beta=\frac{1}{2}$. See the text for more details.}\label{fig2}
\end{figure}

The boundaries of the various solutions are depicted in Fig.~\ref{fig2}. The solid line corresponds to the usual matrix model phase transition, i.e.\ the boundary of the existence domain of the symmetric single-cut solution. It is also the upper-right boundary for the existence domain of the double-cut solution, whose lower-left boundary is the dashed line. The dotted line is the lower-left boundary of the existence domain of the asymmetric single-cut solution. The area in the lower left corner is the forbidden region in parameter space.

\subsection[Solutions of the model on CP**3]{Solutions of the model on $\boldsymbol{\CPP^3}$}

We list now the three solutions for the model on $\CPP^3$. This model dif\/fers from that on $\CPP^2$ only in the magnitude of the coef\/f\/icients in the action. The supports are chosen in the same way as on $\CPP^1$ and $\CPP^2$.

Symmetric single-cut-solution:
\begin{gather*}
 \rho(\lambda)=\frac{3 \beta  \sqrt{d^2-\lambda ^2} \left(10 r-3 (-5+6 c_2 \beta  (1+4 c_2 \beta ))+10 g \left(d^2+2 \lambda ^2\right)\right)}{5 \pi },\\
 c_2=\tfrac{1}{8}(2d^2+d^6g\beta),\qquad
10-3 d^2 \beta  (15 d^2 g+10 r-3 (-5+6 c_2 \beta  (1+4 c_2 \beta )))=0,\\
r>\frac{-5 g \left(9+2 \sqrt{\frac{6g}{\beta }}\right)+16 \beta +6 \sqrt{6 g \beta }}{30 g}.
 \end{gather*}

Double cut solution:
\begin{gather*}
\rho(\lambda)=\frac{2 \lambda  \sqrt{g \beta  \big(6-36 g \beta  \left(s-\lambda ^2\right)^2\big)}}{\pi },\\
c_2=s,\qquad  d=\frac{1}{\sqrt{2\beta g}},\qquad 10 r+20 g s-3 (-5+6 s \beta  (1+4 s \beta ))=0,\\
 \frac{-100 g^2+180 g \beta -1161 \beta ^2}{720 \beta ^2}<r<\frac{-5 g \left(9+2 \sqrt{\frac{6g}{\beta }}\right)+16 \beta +6 \sqrt{6g \beta }}{30 g}.
 \end{gather*}

Asymmetric single-cut solution:
\begin{gather*}
  \rho(\lambda) =\frac{12 g \beta  \sqrt{d^2-(s-\lambda )^2} \left(\lambda  (s+\lambda )-d^2\right)}{\pi } ,\qquad c_2=-\frac{3}{4} d^2 g \left(d^4-10 d^2 s^2-16 s^4\right) \beta ,\\
 s=\frac{\sqrt{2+9 d^4 g \beta }}{2 d \sqrt{6\beta g}}, \qquad 30 d^2 g+10 r+20 g s^2-3 (-5+6 c_2 \beta  (1+4 c_2 \beta ))=0,\\
r>-\frac{3}{2}-\frac{\sqrt{5g}}{\sqrt{2\beta }}+\frac{41 \sqrt{\beta }}{10 \sqrt{10g}}+\frac{1681 \beta }{450 g}.
 \end{gather*}

\begin{figure}[h]
\centerline{\includegraphics{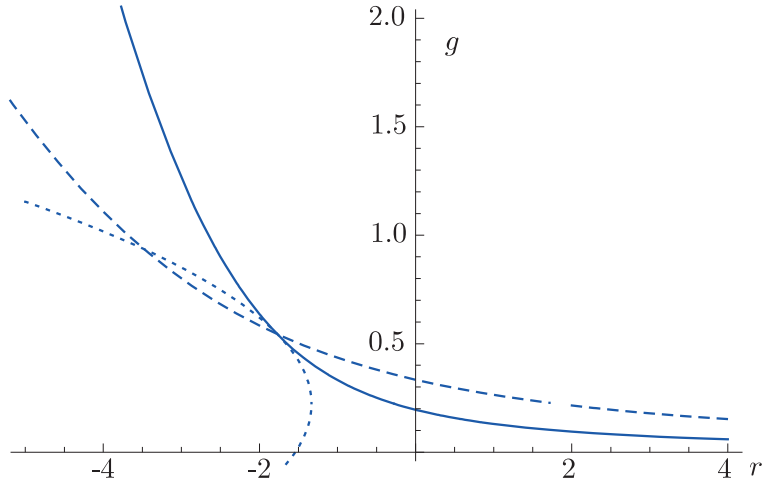}}
\caption{The boundaries of the various existence domains on $\CPP^3$ for $\beta=\frac{1}{2}$.}\label{fig3}
\end{figure}

The boundaries for the existence domains are presented in Fig.~\ref{fig3}. The meaning of the lines is the same as for $\CPP^2$. Not surprisingly, the phase diagram is essentially identical to that for~$\CPP^2$. Unfortunately, there is no feature hinting at the non-renormalizability of $\phi^4$-theory on~$\FR^6$.

\section{Conclusions}\label{section5}

In this paper, we computed the partition function of scalar quantum f\/ield theory on fuzzy $\CPP^n$ to third order in the inverse temperature $\beta$, generalizing the results of \cite{O'Connor:2007ea}. As this theory can be interpreted as a noncommutative deformation of scalar quantum f\/ield theory on $\FR^{2n}$ in the large $N$ limit, we also demonstrated the existence of a nontrivial such theory.

We started by expanding the exponential of the kinetic term in the partition function. We then used group theoretic methods to integrate out the zero modes of the action and obtained a multitrace matrix model. This model was then solved via the saddle point approximation in the large $N$ limit. In principle, however, the partition function of the matrix model could have been computed as well at f\/inite $N$ using orthogonal polynomials.

We presented the explicit classical solutions on which the partition function localizes in the large $N$ limit and discussed the arising phase diagrams for $\CPP^1$, $\CPP^2$ and $\CPP^3$. We conf\/irmed the f\/indings of the numerical analysis of \cite{GarciaFlores:2005xc,Panero:2006bx} for $\CPP^1$ and reproduced qualitatively -- and partly quantitatively -- the phase diagram found numerically. That is, we conf\/irmed the existence of three distinct phases and conf\/irmed analytically their properties suggested by the numerical studies. We also found a triple point which agrees to an acceptable degree with the one found numerically.

Here, it was particularly interesting that we found a large region of the parameter space in which an asymmetric single-cut solution was energetically favorable to a symmetric double-cut solution, even though our potential was symmetric. Such situations have been studied in the past, see e.g.~\cite{Cicuta:1986tm}. Physically, the existence of this spontaneous symmetry breaking in the large~$N$ limit can be explained as follows. Consider the matrix model at f\/inite~$N$. Here, we have to introduce explicitly a symmetry breaking term for an asymmetric phase to exist, as tunnelling of the eigenvalues would otherwise restore symmetry in the eigenvalue f\/illing. After taking $N$ to inf\/inity, there is no more tunnelling, and the symmetry breaking term can be safely switched of\/f, preserving the asymmetric conf\/iguration.

It would be interesting to push the analysis of the phase diagrams further and, for example, to include all odd moments and examine the possibility of a smooth transition of the f\/illing fractions between phases II and III. We then might be able to reproduce the slope for the linear phase boundary found in \cite{GarciaFlores:2005xc}. It might also be interesting to compare our results to the f\/indings of \cite{Gubser:2000cd}, where the phase structure of noncommutative f\/ield theories on Moyal space was analyzed and \cite{Das:2008bc}, where questions arising from \cite{Gubser:2000cd} were discussed on the fuzzy sphere. Moreover, we intend to use the results found here to study scalar f\/ield theory on $\FR\times S^2$ as well as the relation of our multitrace matrix model to (deformed) integrable hierarchies in the future. Finally, recall that multitrace matrix models had been proposed as candidates for conformal f\/ield theories with $c>1$ coupled to gravity \cite{Das:1989fq}. One might be able to make sense of our models in this context, as well.

\vspace{-2pt}

\appendix

\section{Lie algebra conventions}\label{appendixA}

Note that our conventions dif\/fer slightly from those of \cite{O'Connor:2007ea}. Everywhere in our discussion, we use orthonormal hermitian generators $\tau_\mu=\tau_\mu^\dagger$, $\mu=1,\ldots,N^2$ of $\au(N)$, which satisfy\footnote{We always sum over indices which appear twice in a product, irrespective of their positions.}
\begin{equation*}
 \tr(\tau_\mu\tau_\nu)=\delta_{\mu\nu}\qquad \mbox{and}\qquad \tau_\mu^{\alpha\beta}\tau_\mu^{\gamma\delta}=\delta^{\alpha\delta}\delta^{\beta\gamma}.
\end{equation*}
The generators of $\au(N)$ split into the generators $\tau_m$, $m=1,\ldots,N^2-1$ of $\asu(N)$ and $\tau_{N^2}=\frac{1}{\sqrt{N}}\unit_N$. For the former, the Fierz identity reads as
\begin{equation*}
\tau_m^{\alpha\beta}\tau_m^{\gamma\delta}=\delta^{\alpha\delta}\delta^{\beta\gamma}-\frac{1}{N}\delta^{\alpha\beta}\delta^{\gamma\delta}.
\end{equation*}

The Haar measure of $\sSU(N)$ satisf\/ies the following orthogonality relation:
\begin{equation}\label{eq:OrthogonalityRelation}
 \int \dd \mu_H(\Omega)\, [\rho(\Omega)]_{ij} [\rho^\dagger(\Omega)]_{kl} = \frac{1}{\dim(\rho)}\delta_{il}\delta_{jk},
\end{equation}
where $\Omega\in\sSU(N)$, $\rho$ is a f\/inite-dimensional, unitary, irreducible representation and $\rho^\dagger$ denotes its complex conjugate, see e.g.~\cite{O'Connor:2007ea} for the proof.

\section{Projection onto irreducible representations}\label{appendixB}

Given a Young tableau $\lambda$ with rows $\lambda_{i}$ describing a partition of $d$, we def\/ine the symmetrizer $s_\lambda$ and the antisymmetrizer $a_\lambda$ as operations symmetrizing over rows and antisymmetrizing over columns:
\begin{gather*}
 s_\lambda: \ V^{\otimes d}\rightarrow {\rm Sym}^{\lambda_1} V\otimes\cdots\otimes{\rm Sym}^{\lambda_d} V \subset V^{\otimes d},\\
 a_\lambda: \ V^{\otimes d}\rightarrow \Lambda^{\mu_1} V\otimes\cdots\otimes\Lambda^{\mu_d} V \subset V^{\otimes d},
\end{gather*}
where $V$ is the fundamental representation and $\mu_i$ are the rows of the conjugate partition to $\lambda$. The Young symmetrizer $c_\lambda$ is now def\/ined as
\begin{equation*}
 c_\lambda:=s_\lambda a_\lambda,
\end{equation*}
and forms a projectors $c_\lambda^2=\alpha c_\lambda$ with $\alpha$ being the product of all hook lengths of the Young tableau $\lambda$. For more details, see e.g.~\cite{Fulton91representationtheory}. To obtain all irreducible representations, we have to consider all standard Young tableaux, i.e.\ Young tableaux, in which the numbers in the boxes increase both downwards and to the right. At order $\CO(\beta)$, we encounter the tableaux
\begin{equation*}
 \young(1)\otimes\young(2) = \young(12) \oplus \young(1,2)\;,
\end{equation*}
while at order $\CO(\beta^2)$, we need to consider
\begin{gather*}
\young(1)\otimes\young(2)\otimes\young(3)\otimes\young(4)  =  \young(1234) \oplus \young(123,4) \oplus \young(124,3) \oplus \young(134,2)
\oplus \young(12,34) \oplus \young(13,24) \oplus \young(1,2,3,4)\;.
 \end{gather*}

A complication sets in for decomposing tensor products of more than four fundamental representations into irreducible ones in this manner: Young symmetrizers of Young tableaux which correspond to the same Young diagram are no longer mutually orthogonal and have to be orthogonalized by the Gram--Schmidt method. The 76 Young tableaux\footnote{Most of the calculations in the following are done using a computer algebra program.} which originate from the tensor product $\young(1)\otimes\cdots\otimes\young(6)$ correspond to 11 dif\/ferent Young diagrams:
\begin{gather*}
 1:~\tyng(6),\qquad 2:~\tyng(3,3)\,,\qquad 3:~\tyng(4,2)\,,\qquad 4:~\tyng(5,1)\,,\qquad 5:~\tyng(2,2,2)\,,\\
6:~\tyng(3,2,1)\,,\qquad 7:~\tyng(4,1,1)\,,\qquad 8:~\tyng(2,2,1,1)\,,\qquad 9:~\tyng(3,1,1,1)\,,\qquad 10:~\tyng(2,1,1,1,1)\,,\qquad 11:~\tyng(1,1,1,1,1,1)\,.
\end{gather*}
Without spelling them out explicitly, we will order the Young tableaux of a given type lexicographically. That is, a Young tableau precedes another one, if it has a lower number in the f\/irst box in which they dif\/fer (going through the tableaux from left to right and top to bottom). We then label them by their type and their position in the ordering: The tableau $\lambda^{(2,3)}$, for example, is given by the third tableau of the second type:
\begin{equation*}
 \lambda^{(2,3)} = \young(125,346)\,.
\end{equation*}
The projectors $\CP^{(i,j)}_6$ onto the irreducible representations are again given by the Young symmetrizers $c_{\lambda^{(i,j)}}$, except for the ones which are not orthogonal. For example,  $c_{\lambda^{(2,1)}}c_{\lambda^{(2,5)}}\neq 0$ and therefore we need to def\/ine
\begin{equation}\label{eq:modprojs}
 \CP^{(2,1)}_6=\frac{1}{\alpha_2}c_{\lambda^{(2,1)}}-\frac{1}{\alpha_2^2} c_{\lambda^{(2,1)}}c_{\lambda^{(2,5)}},
\end{equation}
where the $\alpha_i$ are the products of the hook lengths of the Young diagram of type $i$, such that $c_{\lambda^{(i,j)}}c_{\lambda^{(i,j)}}=\alpha_ic_{\lambda^{(i,j)}}$. The same is the case for the following pairs of Young symmetrizers:
\begin{alignat*}{6}
 &c_{\lambda^{(3,1)}}c_{\lambda^{(3,8)}},\qquad && c_{\lambda^{(3,2)}}c_{\lambda^{(3,9)}},\qquad && c_{\lambda^{(3,3)}}c_{\lambda^{(3,9)}},\qquad && c_{\lambda^{(5,1)}}c_{\lambda^{(5,5)}},\qquad &&
c_{\lambda^{(6,1)}}c_{\lambda^{(6,11)}},&\\
&c_{\lambda^{(6,1)}}c_{\lambda^{(6,12)}},\qquad && c_{\lambda^{(6,2)}}c_{\lambda^{(6,13)}},\qquad && c_{\lambda^{(6,2)}}c_{\lambda^{(6,14)}},\qquad && c_{\lambda^{(6,3)}}c_{\lambda^{(6,15)}},\qquad && c_{\lambda^{(6,4)}}c_{\lambda^{(6,16)}},& \\
&c_{\lambda^{(6,5)}}c_{\lambda^{(6,15)}},\qquad && c_{\lambda^{(6,7)}}c_{\lambda^{(6,16)}},\qquad &&
c_{\lambda^{(8,1)}}c_{\lambda^{(8,7)}},\qquad && c_{\lambda^{(8,1)}}c_{\lambda^{(8,8)}},\qquad && c_{\lambda^{(8,2)}}c_{\lambda^{(8,9)}}. &
\end{alignat*}
After def\/ining the projectors $\CP^{(i,j)}$ appropriately as in \eqref{eq:modprojs}, they are orthonormal:
\begin{equation*}
 \CP^{(i,j)}_6\CP^{(k,l)}_6=\delta^{ik}\delta^{jl}\CP^{(i,j)}_6.
\end{equation*}
Moreover, the projectors now give the decomposition into irreducible representations: If we have a tuple of 6 objects  $(\alpha_1,\ldots,\alpha_6)$ onto which the Young symmetrizers act, we have
\begin{equation}\label{eq:ProjectorCompleteness}
 (\alpha_1,\ldots,\alpha_6)=\sum_{i,j} \CP^{(i,j)}_6(\alpha_1,\ldots,\alpha_6).
\end{equation}

\section[Detailed results at order O(beta**3)]{Detailed results at order $\boldsymbol{\CO(\beta^3)}$}\label{appendixC}

The contracted restricted trace for $k=3$,
\begin{equation*}
 K_{m_1m_2}\cdots K_{m_5m_6}\tr_{\rho^{(i,j)}}(\tau_{m_1}\otimes\tau_{m_2}\otimes\cdots\otimes\tau_{m_5}\otimes\tau_{m_{6}}),
\end{equation*}
read as:
\begin{alignat*}{3}
\tyng(6)\ :\ &\frac{1}{720 N^2 \tr(K)}\Big(-4 (40+3 N) (1+n) \tr(K)^3+(40+N (12+N)) \tr(K)^4\\
& {}-8 N^3 \tr\left(K^2\right)^2+ 2 N \tr(K)^2 \left(-80 (1+n)^2+(32+3 N) \tr\left(K^2\right)\right)\\
& {}+16 N^2 \tr(K) \left(4 (1+n) \tr\left(K^2\right)+\tr\left(K^3\right)\right)\Big),\\
\tyng(3,3)\ :\ &-\frac{1}{144 N \tr(K)}\Big(-12 (1+n) \tr(K)^3+3 (4+N) \tr(K)^4+16 N^2 \tr\left(K^2\right)^2 \\
&{}+2 \tr(K)^2 \left(16 (1+n)^2+(8+9 N) \tr\left(K^2\right)\right)+24 N \tr(K) \tr\left(K^3\right)\Big),\\
\tyng(4,2)\ :\ &\frac{3}{80 N}\Big(-4 (1+n) \tr(K)^2+(4+N) \tr(K)^3 \\
&{} +2 \tr(K) \left((8+3 N) \tr\left(K^2\right)-16 (1+n)^2\right)+ 8 N \tr\left(K^3\right)\Big),\\
\tyng(5,1)\ :\ &\frac{1}{144 N^2 \tr(K)}\Big(4 (40+3 N) (1+n) \tr(K)^3-(40+N (12+N)) \tr(K)^4\Big.\\
&{}+8 N^3 \tr\left(K^2\right)^2+2 N \tr(K)^2 \left(80 (1+n)^2-(32+3 N) \tr\left(K^2\right)\right)\\
&{} -16 N^2 \tr(K) \left(4 (1+n) \tr\left(K^2\right)+\tr\left(K^3\right)\right)\Big),\\
\tyng(2,2,2)\ :\ &\frac{1}{144 N \tr(K)}\Big(12 (1+n) \tr(K)^3+3 (-4+N) \tr(K)^4-16 N^2 \tr\left(K^2\right)^2 \\
&{}-2 \tr(K)^2 \left(16 (1+n)^2+(8-9 N) \tr\left(K^2\right)\right)+24 N \tr(K) \tr\left(K^3\right)\Big),\\
\tyng(3,2,1)\ :\ &\frac{16 \left(2 \tr(K)^2 \left(4 (1+n)^2-\tr\left(K^2\right)\right)+N^2 \tr\left(K^2\right)^2\right)}{45 N \tr(K)},
 \\
\tyng(4,1,1)\ :\ &-\frac{1}{36 N^2 \tr(K)}\Big(80 (1+n) \tr(K)^3+\left(-20+N^2\right) \tr(K)^4+4 N^3 \tr\left(K^2\right)^2 \\
& {}+2 N \tr(K)^2 \left(16 (1+n)^2+(-4+3 N) \tr\left(K^2\right)\right)\\
&{} +4 N^2 \tr(K) \left(-8 (1+n) \tr\left(K^2\right)+\tr\left(K^3\right)\right)\Big),\\
\tyng(2,2,1,1)\ :\ &-\frac{3}{80 N} \Big(4 (1+n) \tr(K)^2+(-4+N) \tr(K)^3\\
&{}+2 \tr(K) \left(16 (1+n)^2+(-8+3 N) \tr\left(K^2\right)\right)+ 8 N \tr\left(K^3\right)\Big),\\
\tyng(3,1,1,1)\ :\ & \frac{1}{36 N^2 \tr(K)}\Big(80 (1+n) \tr(K)^3+\left(-20+N^2\right) \tr(K)^4-4 N^3 \tr\left(K^2\right)^2 \\
&{}+2 N \tr(K)^2 \left(-16 (1+n)^2+(4+3 N) \tr\left(K^2\right)\right)\\
&{} +4 N^2 \tr(K) \left(-8 (1+n) \tr\left(K^2\right)+\tr\left(K^3\right)\right)\Big),\\
\tyng(2,1,1,1,1)\ :\ &\frac{1}{144 N^2 \tr(K)}\Big(4 (-40+3 N) (1+n) \tr(K)^3+(40+(-12+N) N) \tr(K)^4 \\
& +8 N^3 \tr\left(K^2\right)^2+ 2 N \tr(K)^2 \left(80 (1+n)^2+(-32+3 N) \tr\left(K^2\right)\right)\\
&{} +16 N^2 \tr(K) \left(4 (1+n) \tr\left(K^2\right)+\tr\left(K^3\right)\right)\Big),\\
\tyng(1,1,1,1,1,1)\ :\ &-\frac{1}{720 N^2 \tr(K)}\Big(4 (-40+3 N) (1+n) \tr(K)^3+(40+(-12+N) N) \tr(K)^4 \\
& {} +8 N^3 \tr\left(K^2\right)^2+ 2 N \tr(K)^2 \left(80 (1+n)^2+(-32+3 N) \tr\left(K^2\right)\right)\\
&{}+16 N^2 \tr(K) \left(4 (1+n) \tr\left(K^2\right)+\tr\left(K^3\right)\right)\Big).
 \end{alignat*}

In the following, we present some more details needed in the calculation of the contracted traces. We start by computing the following traces over the generators of $\asu(n+1)$ in the $N$-dimensional representation:
\begin{gather*}
  \tr(L_iL_jL_iL_j)=\frac{\tr(K)^2}{4N^3}-(n+1)\frac{\tr(K)}{2N},\\
  \tr(L_iL_jL_kL_iL_jL_k)=\frac{\tr(K)(8(1+n)^2N^4-6(1+n)N^2\tr(K)+\tr(K)^2}{8N^5},\\
  \tr(L_iL_jL_kL_jL_iL_k)=\frac{\tr(K)(-2(1+n)N^2+\tr(K))^2}{8N^5},\\
  \tr(L_iL_jL_k)\tr(L_jL_iL_k)=-\frac{1}{8}\tr(K^3)-\frac{\tr(K)^3}{4N^4}+\frac{3\tr(K)\tr(K^2)}{8N^2},\\
  \tr(L_iL_jL_k)\tr(L_iL_jL_k)=\tr(L_iL_jL_k)\tr(L_jL_iL_k)+\frac{(1+n)(\tr(K)^2-N^2\tr(K^2))}{4N^2}.
 \end{gather*}
Using these, we obtain
\begin{gather*}
K_{\mu_1\nu_1}K_{\mu_2\nu_2}\tr(\tau_{\mu_1}\tau_{\nu_1}\tau_{\mu_2}\tau_{\nu_2})= \frac{\tr(K)^2}{N},\\
K_{\mu_1\nu_1}K_{\mu_2\nu_2}\tr(\tau_{\mu_1}\tau_{\mu_2}\tau_{\nu_1}\tau_{\nu_2})= 2\tr([L_i,L_j][L_i,L_j])=-2(n+1)\frac{\tr(K)}{N},\\
K_{\mu_1\nu_1}K_{\mu_2\nu_2}K_{\mu_3\nu_3}\tr(\tau_{\mu_1}\tau_{\nu_1}\tau_{\mu_2}\tau_{\nu_2}\tau_{\mu_3}\tau_{\nu_3})= \frac{\tr(K)^3}{N^2},\\
K_{\mu_1\nu_1}K_{\mu_2\nu_2}K_{\mu_3\nu_3}\tr(\tau_{\mu_1}\tau_{\nu_1}\tau_{\mu_2}\tau_{\mu_3}\tau_{\nu_2}\tau_{\nu_3})= -\frac{2(n+1)\tr(K)^2}{N^2},\\
K_{\mu_1\nu_1}K_{\mu_2\nu_2}K_{\mu_3\nu_3}\tr(\tau_{\mu_1}\tau_{\nu_1}\tau_{\mu_2}\tau_{\mu_3}\tau_{\nu_3}\tau_{\nu_2})= \frac{\tr(K)^3}{N^2},
\\
K_{\mu_1\nu_1}K_{\mu_2\nu_2}K_{\mu_3\nu_3}\tr(\tau_{\mu_1}\tau_{\mu_2}\tau_{\nu_1}\tau_{\mu_3}\tau_{\nu_2}\tau_{\nu_3})= 2(n+1)\left(\tr(K^2)-\frac{\tr(K)^2}{N^2}\right),\\
K_{\mu_1\nu_1}K_{\mu_2\nu_2}K_{\mu_3\nu_3}\tr(\tau_{\mu_1}\tau_{\mu_2}\tau_{\mu_3}\tau_{\nu_1}\tau_{\nu_2}\tau_{\nu_3})= \tr(K^3)+2(n+1)\left(\tr(K^2)-\frac{\tr(K)^2}{N^2}\right),
\end{gather*}
as well as
\begin{gather*}
  K_{\mu_1\nu_1}K_{\mu_2\nu_2}K_{\mu_3\nu_3}\tr(\tau_{\mu_1}\tau_{\mu_2}\tau_{\mu_3})\tr(\tau_{\nu_1}\tau_{\nu_2}\tau_{\nu_3})=-\frac{8(1+n)^2\tr(K)}{N},\\
  K_{\mu_1\nu_1}K_{\mu_2\nu_2}K_{\mu_3\nu_3}\tr(\tau_{\mu_1}\tau_{\mu_2}\tau_{\mu_3})\tr(\tau_{\nu_1}\tau_{\nu_3}\tau_{\nu_2})=\frac{2 \tr(K) \tr\left(K^2\right)}{N}-\frac{N \tr\left(K^2\right)^2}{\tr(K)},\\
K_{\mu_1\nu_1}K_{\mu_2\nu_2}K_{\mu_3\nu_3}\tr(\tau_{\mu_1}\tau_{\mu_2})\tr(\tau_{\nu_1}\tau_{\nu_2}\tau_{\mu_3}\tau_{\nu_3})=\frac{\tr(K) \tr\left(K^2\right)}{N},\\
K_{\mu_1\nu_1}K_{\mu_2\nu_2}K_{\mu_3\nu_3}\tr(\tau_{\mu_1}\tau_{\mu_2})\tr(\tau_{\nu_1}\tau_{\mu_3}\tau_{\nu_2}\tau_{\nu_3})=-\frac{4(1+n)^2\tr(K)}{N}.
\end{gather*}
These expressions can be easily checked for $n=1$ and $N=2,3,4$ by comparing them to the results of a direct computation.

\subsection*{Acknowledgements}

I would like to thank Denjoe O'Connor, Richard Szabo and Miguel Tierz for helpful discussions and comments on a draft of this paper. I am also grateful to Denjoe O'Connor for bringing this problem to my attention two years ago and for collaborating on the previous paper~\cite{O'Connor:2007ea}. This work was supported by a Career Acceleration Fellowship from the UK Engineering and Physical Sciences Research Council.

\pdfbookmark[1]{References}{ref}
\LastPageEnding

\end{document}